# Design, Status and Physics Potential of JUNO

**Hans Th. J. Steiger** [a,b,*,1]

[a] *Cluster of Excellence PRISMA⁺,*
*Staudinger Weg 9, 55128 Mainz, Germany*

[b] *Institute of Physics, Johannes Gutenberg University Mainz,*
*Staudinger Weg 7, 55128 Mainz, Germany*

*E-mail:* `hsteiger@uni-mainz.de`

The Jiangmen Underground Neutrino Observatory (JUNO) is a 20 kton multi-purpose liquid scintillator detector currently being built in a dedicated underground laboratory in Jiangmen (PR China). Data-taking is expected to start in 2023. JUNO's main physics goal is the determination of the neutrino mass ordering using electron anti-neutrinos from two nuclear power plants at a baseline of about 53 km. JUNO aims for an unprecedented energy resolution of 3 % at 1 MeV for the central detector, which will allow determining the mass ordering with 3 $\sigma$ significance within six years of operation. Furthermore, measurements in neutrino physics and astrophysics, such as estimating the solar oscillation parameters and the atmospheric mass splitting with an accuracy of 0.5 % or better, will be performed. In these proceedings, JUNO's design, the status of its construction, and its physics potential, will be presented alongside a short excursion into its rich R&D program.

---

[*]Speaker
[1] on behalf of the JUNO Collaboration

## 1. Introduction

The Jiangmen Underground Neutrino Observatory (JUNO) [1] is a multi-purpose neutrino observatory but is focused mainly on the determination of the neutrino mass ordering (NMO) as primary physics goal [2]. Currently, the unknown sign of $\Delta m^2$ leads to two possible scenarios, called the "normal ordering" (NO) where $m_1 < m_2 < m_3$ and the "inverted ordering" (IO) with $m_3 < m_1 < m_2$. As the value of $\theta_{13}$ is relatively large, a medium baseline reactor antineutrino oscillation experiment like JUNO provides an excellent opportunity to resolve the NMO realized in nature. While experiments based on accelerators (e.g. DUNE [3], NOvA [4]) or atmospheric neutrinos (e.g. ORCA [5], HyperK [6]) exploit the matter effect of neutrino oscillations, JUNO is unique as it will perform the NMO determination through the simultaneous observation of $\Delta m^2_{31}$ and $\Delta m^2_{32}$. In fact, this leads to the advantage, that JUNO's NMO sensitivity is independent from the CP-violating phase and the octant of $\theta_{23}$, especially when combined with other experiments. The survival probability for the reactor antineutrinos in vacuum can be expressed as

$$P_{\bar{\nu}_e \to \bar{\nu}_e} = 1 - sin^2 2\theta_{13}(cos^2\theta_{12} sin^2\Delta_{31} + sin^2\theta_{12} sin^2\Delta_{32}) - cos^4\theta_{13} sin^2 2\theta_{12} sin^2\Delta_{21}$$

where $L$ denotes the baseline and $\Delta_{ij} = \Delta m^2_{ij} L/(4E)$. The expected antineutrino energy spectrum for the JUNO baseline of ∼ 53 km is shown in Figure 1.

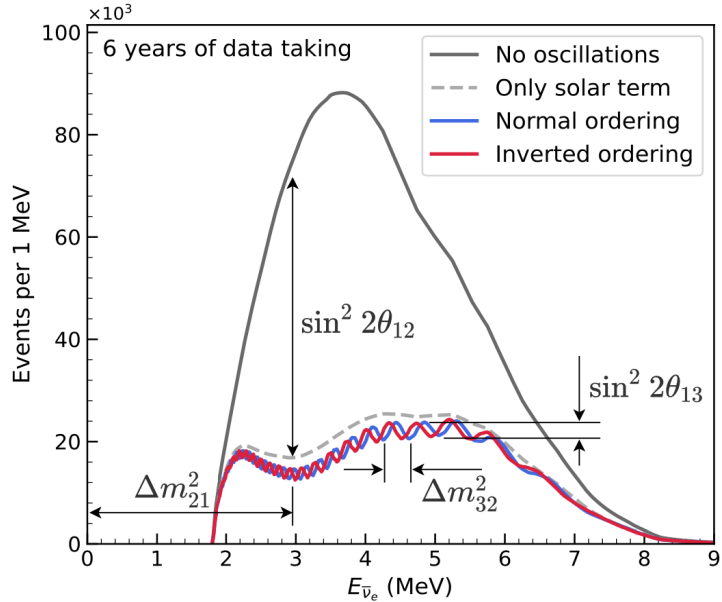

**Figure 1:** Calculated energy spectrum for the JUNO site. Note that JUNO will be able to simultaneously measure oscillations caused by the small mass splitting ($\Delta m^2_{21}$) and large mass splitting ($\Delta m^2_{31}$ and $\Delta m^2_{32}$).

To be able to distinguish the NO from the IO pattern, a large neutrino target mass of JUNO (20 kt) and an unprecedented energy resolution ($\sigma_E/E$) of 3 % at 1 MeV of the detector are key ingredients for a NMO determination with 3 σ significance within six years.

Due to these strict requirements in detector performance, JUNO will be able to measure most neutrino oscillation parameters in the solar and atmospheric sectors with an accuracy of 1 % or better. Furthermore, JUNO will measure geo-neutrinos, contribute to the search for the Diffuse Supernova Neutrino Background (DSNB) and it will monitor the neutrino sky continuously for contributing to neutrino and multi-messenger astronomy. Beside this, JUNO will perform a proton decay search in the channel $p \to K^+ \bar{\nu}$. A complete description of JUNO's physics program can be found in [1, 2].

## 2. The JUNO Experiment

The JUNO site is located in Jinji town, about 43 km South-West of Kaiping city, in Guangdong province, China. Two nuclear power plants Yangjiang and Taishan are located in an equal distance of ∼ 53 km to the JUNO central detector. A nominal power of 26.6 GWth from both power stations will be available. The main detector is being constructed in a dedicated underground laboratory below Dashi hill, resulting in an overburden of about 700 m solid rock (∼ 1800 m.w.e.). Figure 2 gives an overview of the central detector's design.

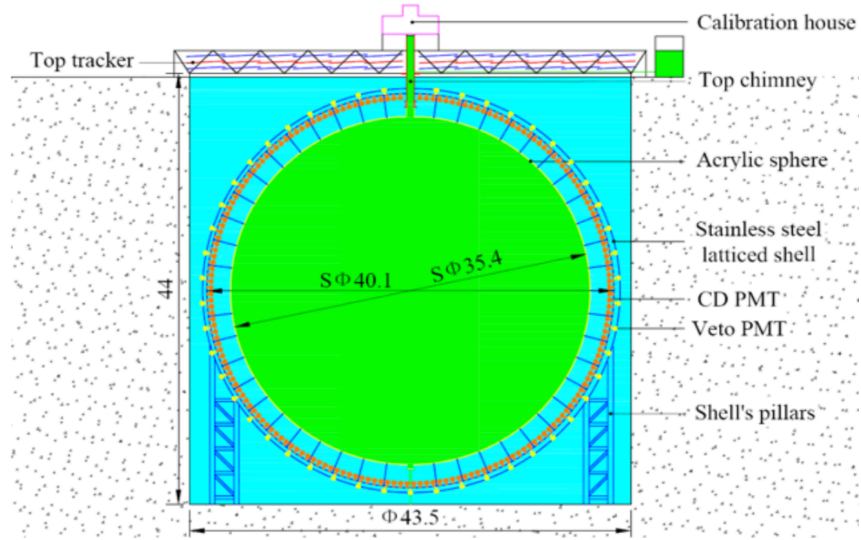

**Figure 2:** Simplified scheme of the JUNO central detector. A 12 cm thick acrylic hollow sphere will contain the 20 kt LS neutrino target. A surrounding water pool is instrumented as muon veto system.

The JUNO main detector consists of the Central Detector (CD), a water Cherenkov detector surrounding it the Top Tracker (TT) above the center of the water pool. The CD, a spherical Acrylic Vessel with inner diameter of 35.4 m and 12 cm thickness is supported by a stainless steel shell structure (SS) and contains 20 kton of LS. The SS supports 17,612 20-inch PMTs and 25,600 3-inch PMTs reading out the scintillaton light. Additional 2,400 20-inch PMTs instrument the water volume to make it a water Cherenkov detector acting as veto system. The CD and the water volume are optically separated. The front-end electronics with its cabling is also directly mounted at the stainless steel structure as well as the magnetic-coils designed to compensate the geomagnetic field.

### 2.1 The 20-inch PMTs

Due to the demanding energy resolution required by the physics program of JUNO, one of the key components is the photon detection system to maximize the collection of the scintillation light. The total photocathode coverage of the detector will be around ∼78 %, where approx. ∼ 75 % are provided by the 20-inch PMT system. JUNO's inventory of these large tubes consists of 5,000 dynode PMTs (R12860HQE) produced by Hamamatsu Photonics K.K., while the remaining PMTs are Micro Channel Plate PMTs manufactured by North Night Vision Technology Co. Ltd. (NNVT).

#### 2.1.1 The PMT-testing facility

While the small PMTs are tested by HZC in corporation with the JUNO Collaboration at the factory site, the large PMTs are stored and tested in Zhongshan Pan-Asia. There dedicated performance testing systems were installed and run since July 2017 [7]. Two container test systems with electromagnetic shielding and commercial electronics were designed for mass acceptance tests. Each testing container houses 36 individual drawers (see Fig. 3 a and b).

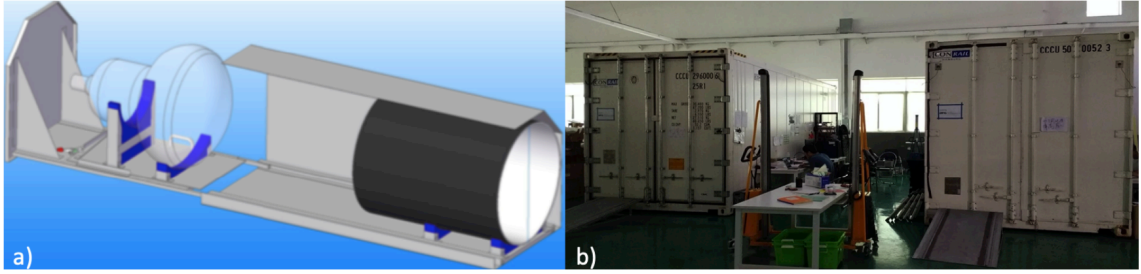

**Figure 3:** a) One of the PMT drawers as CAD Model. b) The two PMT testing containers in Zhongshan Pan-Asia.

The PMTs can be pulsed with a self-stabilized LED or a picosecond laser, both with 420 nm wavelength. A commercial switched-capacitor ADC (CAEN V1742) is used for waveform readout. Triggering and noise counting is realized via VME based leading edge discriminators (CAEN V895B) and latching scalers (CAEN V895AC). Using a PCIe computer bridge card and a custom made LabView based software, the entire system is controlled.

In the containers the parameters charge resolution, single PE peak/valley ratio, operating voltage necessary for a gain of $10^7$, dark count rate (DCR), single PE rise and fall time as well as pre pulsing and after pulsing rate are determined. The PMTs from the two companies are well comparable in performances, and it should be mentioned that the PMT's detection efficiency (quantum efficiency × collection efficiency) has reached a value close to 30% [8]. The testing results show that the NNVT PMTs have lower after pulse probability and lower radioactive background, while the Hamamatsu PMTs have lower transit time spread.

Beside the containers, two scanning stations were set up to irradiate the photo cathodes from 168 point-like sources. The scanning provides information about the non-uniformity of PMT parameters, as well as their earth magnetic field dependence (for a detailed description see [9])

## 2.2 The Liquid Scintillator

The LS mixture consists of three components: LAB as a solvent, 2.5 g/l PPO (2,5-Diphenyl-oxazole) acting as fluor and 1-3 mg/l bis-MSB (1,4-bis-(o-methylstyryl)-benzene) as secondary wavelength shifter. The optical requirements are a light yield of ∼ $10^4$ photons per MeV energy deposition and an attenuation length for 430 nm light above 20 m. For reactor neutrinos, mass concentrations of $^{238}U$ and $^{232}Th$ below $10^{-15}$ g/g and for $^{40}K$ $10^{-16}$ g/g are required. For effective solar neutrino spectroscopy these concentrations should be two orders of magnitude lower, while the amount of $^{14}C$ should not exceed $10^{-18}$ g/g. To achieve these values, the LS will be purified in a series of dedicated process. In a first step the LAB will be filtered in an $Al_2O_3$ column. After that, it will be distilled to further improve the transparency and remove heavy radioactive metals. In the following $^{40}K$ and more isotopes from the uranium and thorium chain will be removed by water extraction. Gaseous impurities like argon, krypton and radon are removed by steam stripping. A detailed description of distillation and stripping systems and first results from a pilot plant test phase at the Daya Bay detector site can be found in [10, 11].

## 2.3 The OSIRIS Pre-Detector Facility

The Online Scintillator Internal Radioactivity Investigation System (OSIRIS) [12] is currently constructed as a failsafe monitor to aid the commissioning of the on-site scintillator purification plants and assess the quality of the scintillator batches in terms of radioactivity before filling them into the central detector of JUNO. OSIRIS will be a liquid scintillator detector (∼19 t LS target) determining uranium and thorium concentrations via the fast Bi/Po-coincidence signals of the corresponding decay-chains. Other contaminations (e.g. $^{14}C$, $^{85}K$) will also be evaluated. OSIRIS contains an acrylic cylinder filled with LS, with a dimension of 3 m in diameter and 3 m in height, placed in the center of a cylindrical water tank of 10 m diameter and 11 m height. The vessel will be instrumented with 64 PMTs (R12860HQE) to detect the scintillation light. The external tank, filled with ultra-pure water, will also be equipped with 12 PMTs (R12860HQE) to allow an operation as a water Cherenkov veto detector in order to reduce the background from crossing muons as well as radioactivity in surrounding rocks and the detector's PMTs [12, 13]. According to Monte Carlo studies radioactivity measurements down to the level necessary for the JUNO IBD requirements will be possible within ∼ 24 hours, while the solar limits could be reached within one week [12].

After the completion of the OSIRIS program, an upgrade of the detector is planned. This upgraded facility, named Serappis (SEarch for RAre PP-neutrinos In Scintillator) aims at a precision measurement of the flux of solar pp-neutrinos on the few-percent level. Such a measurement will be a relevant contribution to the study of solar neutrino oscillation parameters and a sensitive test of the solar luminosity constraint. To go substantially beyond current accuracy levels for the pp-flux, an organic scintillator with ultra-low $^{14}$C levels (below $10^{-18}$) is required. The already existing OSIRIS detector and JUNO infrastructure will be used for the identifying of suitable scintillator materials. By Serappis a unique chance for a low-budget, high-precision studies of fundamental properties of our sun will be possible [14].

## 3. The Taishan Antineutrino Observatory (TAO)

As the present information on the reactor neutrino spectra is not meeting the requirements of an experiment like JUNO, with a design resolution of 3 % at 1 MeV, the JUNO Collaboration is currently constructing the Taishan Antineutrino Observatory (TAO) as a high resolution near detector facility. The TAO apparatus (see Fig. 4) will be installed at a distance of 30 m from the 4.6 GW$_{th}$-power core of the Taishan-1 plant. Unknown fine structures in the reactor spectrum might cause severe uncertainties, which could even make the interpretation of JUNO's reactor neutrino data impossible. TAO is aiming for a measurement of the reactor neutrino spectrum with a groundbreaking resolution better than ∼ 2 % at 1 MeV [15]. Furthermore, TAO will make a major contribution in the investigation of the so-called reactor anomaly. Present calculations of the reactor neutrino spectrum indicate a deficit of approximately 3 % in the measured reactor fluxes. Currently, these anomalies can be interpreted as indications for the existence of right-handed sterile neutrinos.

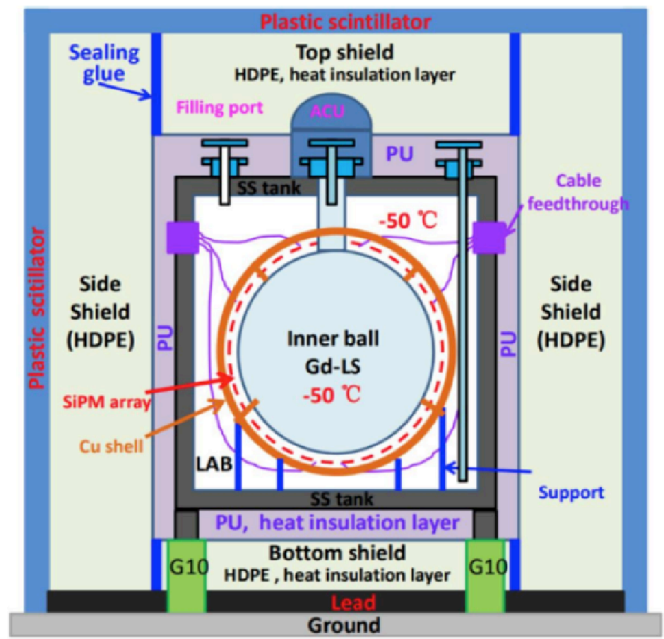

**Figure 4:** Schematic view of the TAO detector.

The TAO experiment will realize the unprecedented neutrino detection rate of about ∼ 2000 per day, which is approximately ∼ 36 times the rate in the JUNO main detector. In order to achieve its goals, TAO is relying on cutting-edge technology, both in photosensor and liquid scintillator development. The experiment will realize an optical coverage of the 2.6 tons of Gd-loaded LS close to ∼ 95 % with novel silicon photomultipliers (SiPMs), with a photon detection efficiency of ∼ 50 %. To efficiently reduce the dark count of these light sensors, the entire detector will be cooled down to -50°C. The combination of SiPMs with cold LS will lead to an increase in the photo electron yield by a factor of ∼ 4.5 compared to the JUNO central detector [15].

## 4.　　Summary and Outlook

When completed JUNO will be the largest LS detector ever built. When the energy resolution of 3 % at 1 MeV can be realized, JUNO can be expected to resolve the neutrino mass ordering at 3 σ within 6 years of data taking. Moreover, many other important topics in neutrino- and particle-physics will be investigated. The detector design has been finalized and all challenges regarding the detector technologies have been solved. The detector component production and installation at the JUNO site is ongoing. Detector construction is expected to complete by the end of 2022.

**References**

[1] A. Abusleme et al., *JUNO Physics and Detector*, A. Abusleme et al, Progr. Part. Nucl. Phys. 123, (2022), 103927.

[2] F. An et al., *Neutrino Physics with JUNO*, J. Phys. **G 43** (3) (2016) 030401, arXiv:1507.05613.

[3] B. Abi, et al., *Long-baseline neutrino oscillation physics potential of the DUNE experiment*, arXiv:2006.16043.

[4] D. S.' Ayres et al., *NOvA: Proposal to Build a 30 Kiloton Off-Axis Detector to Study $\nu_\mu \rightarrow \nu_e$ Oscillations in the NuMI Beamline*, (2004). arXiv:hep-ex/0503053.

[5] P. Kalaczyński, *KM3NeT/ORCA: status and perspectives for neutrino oscillation and mass hierarchy measurements*, PoS (ICHEP2020) 149, arXiv:2107.10593.

[6] K. Abe et al., *Letter of Intent: The Hyper-Kamiokande Experiment — Detector Design and Physics Potential*, (2011), arXiv:1109.3262.

[7] JUNO collaboration, *Large photocathode 20-inch PMT testing methods for the JUNO experiment*, (2017), JINST 12 C06017, arXiv:1705.05012.

[8] Zhang, Z. Wang, F. Luo, A. Yang, Z.-H. Qin, C. Yang et al., *Study on relative collection efficiency of PMTs with spotlight*, Radiat. Detect. Technol. Meth. 3 (2019) 20.

[9] Anfimov N., *Large photocathode 20-inch PMT testing methods for the JUNO experiment*, JINST, 2017, 12(06): C06017.

[10] P. Lombardi et al., *Distillation and stripping pilot plants for the JUNO neutrino detector: Design, operations and reliability,* NIM-A, 925 , 6-17, 2019.

[11] The JUNO Collaboration, *Optimization of the JUNO liquid scintillator composition using a Daya Bay antineutrino detector,* NIM-A, 988, 164823, arXiv:2007.00314, 2021.

[12] The JUNO Collaboration, *The design and sensitivity of JUNO's scintillator radio purity pre-detector OSIRIS*, accepted by Eur. Phys. J. C, arXiv:2103.16900, 2021.

[13] H. Steiger, *OSIRIS: Status of the Detector Hardware Development*, https://indico.cern.ch/event/738555/contributions/3174135/attachments/1736380/2808991/OSIRIS_Hardware-Finland18.pdf.

[14] L. Bieger et al., *Potential for a precision measurement of solar pp neutrinos in the Serappis Experiment,* accepted by Eur. Phys. J. C, arXiv: 2109.10782, 2021.

[15] JUNO Collaboration, *TAO Conceptual Design Report,* arXiv: 2005.08745.